# A Software-Defined Networking Solution for Interconnecting Network Functions in Service-Based Architectures

PABLO FONDO-FERREIRO, FELIPE GIL-CASTIÑEIRA,
FRANCISCO JAVIER GONZÁLEZ-CASTAÑO, AND DAVID CANDAL-VENTUREIRA
Information Technologies Group, AtlanTTic, University of Vigo, 36310 Vigo, Spain

Corresponding author: Felipe Gil-Castiñeira (xil@gti.uvigo.es)

This work was supported in part by the ''la Caixa'' Foundation (ID 100010434) Fellowship under Grant LCF/BQ/ES18/11670020; and in part by the Ministerio de Ciencia e Innovación, Spain, under Grant PID2020-116329GB-C21 and Grant PDC2021-121335-C21.

**ABSTRACT** Mobile core networks handle critical control functions for delivering services in modern cellular networks. Traditional point-to-point architectures, where network functions are directly connected through standardized interfaces, are being substituted by service-based architectures (SBAs), where core functionalities are finer-grained microservices decoupled from the underlying infrastructure. In this way, network functions and services can be distributed, with scaling and fail-over mechanisms, and can be dynamically deployed, updated, or removed to support slicing. A myriad of network functions can be deployed or removed according to traffic flows, thereby increasing the complexity of connection management. In this context, 3GPP Release 16 defines the service communication proxy (SCP) as a unified communication interface for a set of network functions. In this paper, we propose a novel software-defined networking (SDN)-based solution with the same role for a service mesh architecture where network functions can be deployed anywhere in the infrastructure. We demonstrated its efficiency in comparison with alternative architectures.

## I. INTRODUCTION
### A. MOTIVATION

5G networks are expected to cope not only with a large amount of traffic but also with a huge number of devices (e.g., IoT), which will challenge the control plane [1]. For example, massive machine-type communications (mMTC) will introduce considerable signaling load in the control plane of the 5G core network (5GC) [2]. The 5GC is defined as a service-based architecture (SBA) whose network functions (NFs) expose their functionality as services, allowing other authorized NFs to access them [3]. Nowadays, web-level technologies are usually adopted to implement the 5GC and address challenges such as congestion control, traffic prioritization, overload control, and optimized routing [4]. Hence, microservice architectures can be used for its implementation, as a service mesh with plenty of functions that can be dynamically instantiated or removed. This allows adapting the control plane to the traffic load and to the requirements of the operator. However, this flexibility comes at a cost. For example, NFs must discover each other to find those that offer the required services and select the most adequate ones in terms of load, network topology, etc.

To overcome the challenges of the SBA approach, the 3rd Generation Partnership Project (3GPP) recently standardized in 5G Release 16 an optional entity for the 5GC SBA called service communication proxy (SCP). The SCP can be seen as a middleware for handling signaling communications among the NF services in the 5GC. The SCP provides indirect communication between NFs, delegated discovery of NF services, load balancing, and traffic monitoring. It allows NF services to focus on their business logic, by assuming inter-service communication tasks [5].

However, NFs must be explicitly designed and configured to use the SCP (e.g., configuration of the SCP endpoint in the NFs, use of SCP-specific parameters in the signaling

The associate editor coordinating the review of this manuscript and approving it for publication was Tiankui Zhang.

communications) and therefore the SCP entity is not transparent to the NFs. It also increases the complexity of the core network since an additional entity is being introduced in the architecture. Besides, it may impose communications overhead to the NF services of the 5G core network because an additional entity will process 5GC signaling traffic. Hence, it is necessary to understand how the different SCP deployment proposals may affect the performance of the 5GC and study alternatives that could lead to better results.

### B. CONTRIBUTIONS

In this work, we analyze how SDN can be leveraged to provide the desired SCP functionalities both transparently to the NFs and in a standard-compliant manner. We propose a novel SDN-based solution as an alternative to the SCP, in line with the trend of SDN as a key enabler for 5G networks and future mobile networks [6], [7]. Our proposal can also be understood as a mechanism for implementing the SCP functionalities in the 5GC transparently to the NFs. In this regard, this work is an extension to the 5GC SBA defined by the 3GPP.

In detail, our main contributions are:

- A novel solution based on SDN for interconnecting NFs in the 5GC that is an alternative to the SCP. This solution addresses the challenges of service-based architectures (e.g. 5G-aware load balancing, traffic forwarding, etc). The proposed SDN mechanism is completely transparent because NFs do not need to include additional parameters in service requests/responses, as the SDN controller handles the policies internally. It does not increase the complexity of the network because communications are directly handled through SDN without any intermediary agents. In addition, it reduces the communications overhead introduced by SCP agents since SDN switches can process most packets at line rate. This is achieved by moving the decision logic from the SCP agents to the SDN controller.
- An experimental performance analysis of the proposed solution.
- A comparison of the proposed solution with agent-based SCP deployments proposed by 3GPP.

The remainder of this paper is organized as follows: Section II discusses the background and related work, and Section III describes the proposed solution based on SDN. Section IV analyzes and compares it with alternative SCP deployments. Finally, Section V concludes the paper.

## II. BACKGROUND AND RELATED WORK
### A. 5G CORE NETWORK SERVICE-BASED ARCHITECTURE

The 5GC is composed of a set of NFs that expose their functionalities as services, so that it is possible to orchestrate, scale and re-use them in a distributed environment [3]. An NF service is a type of capability that an NF (the NF service producer) is exposed to other authorized NF (the NF service consumer) through a service-based interface (SBI). In this architecture, system procedures are described as a sequence of NF service invocations.

A service framework should provide the following essential functionalities: service registration/de-registration, consumer authorization, service discovery, and inter-service communication. The NF repository function (NRF) is a key NF within the service framework. It supports service registration and discovery by maintaining the NF profile of available NFs and their services. The NRF also notifies the registration, updates, and de-registration of the NF instances. NF service consumers use the service framework to contact NF service producers. First, the framework must discover and select the target NF service producer before sending the message.

An example of communication between NFs takes place during a user equipment (UE) session establishment, where the access and mobility management function (AMF) needs to communicate with the session management function (SMF) to create the corresponding session context. For this purpose, the AMF first performs a service discovery through the `Nnrf_NFDiscovery` service of the NRF to obtain a list of available SMFs in the network. In this communication, the AMF acts as a consumer, while the NRF acts as a producer. After obtaining the list of available SMFs, the AMF selects one of them and starts the establishment of the session for the UE: the AMF, acting as a consumer, invokes the `CreateSMContext` operation of the `Nsmf_PDUSession` service exposed by the SMF, which acts as a producer. For the discovery step to be successful, some SMF should have been previously registered using the `NFRegister` operation of the `Nnrf_NFManagement` service provided by the NRF. This registration procedure is typically executed at startup when the NF is deployed. Note that these three communication examples follow the request/response semantics. The procedures defined for the 5GC are described in 3GPP TS 23.502 [8].

As a microservice architecture, the NFs in the 5G SBA are expected to be implemented using lightweight containers as the underlying virtualization technology. Therefore, container orchestration tools, such as Kubernetes, become relevant. This is also the case for emergent open-source service mesh solutions, such as Linkerd, Consul, and Istio. However, generic service mesh solutions have some limitations for the realization of the 5G SBA, as identified in [4]: they are unable to leverage 5G-specific information for network configurations (e.g., the NF profile information stored in the NRF for making load-balancing decisions).

The 3GPP Release 16 standard considers two types of communication between an NF service consumer and an NF service producer: direct communication and indirect communication (via the SCP).

- Direct communication: The NF service consumer performs discovery by local configuration or via the NRF. The NF service consumer communicates directly with the NF service producer.

- Indirect communication: The NF service consumer communicates with the NF service producer via the SCP. The NF service consumer may perform the discovery directly or delegate it to the SCP.

The direct communication option might seem the best choice since it is simpler and does not introduce overhead. However, it may not guarantee the correct operation of the network in complex scenarios. As noted before, the SCP was designed to collaborate with the NRF to allow elastic growth, interoperability, and rapid introduction of new services [4].

The protocols used by the SBA allow for two types of interaction mechanisms between NF service consumers and producers: request-response (synchronous communication) and subscribe-notify (asynchronous communication). The 3GPP has selected HTTP as the protocol for the SBI, as indicated in 3GPP TS 29.500 [9].

### B. SERVICE COMMUNICATION PROXY

3GPP introduced the SCP as a 5GC entity in Technical Specification (TS) 23.501 Release 16 [3]. It provides the following functionalities (see Section 6.2.19 of 3GPP TS 23.501 [3]):

- Indirect NF/NF service communications.
- Delegated discovery in NF/NF service (e.g. the NF consumer does not look for the NF producer in the NRF, but instead delegates the discovery to the SCP).
- Message forwarding and routing to destination in NF/NF service.
- Communications security (e.g. authorization of the NF service consumer to access the NF service producer API).
- Load balancing, monitoring, and overload control.
- Routing binding indication: Ensures that the same NF/NF service instances are used in subsequent communications.

As previously mentioned, the SCP hides the internal dynamics of the 5GC network to the NFs, so that each NF can focus on its own functionality and use the SCP as a message-passing tool. However, the SCP is an optional entity in the 5GC, and NFs can be configured to establish direct communications.

Thus, the SCP can be seen as a middleware layer that allows NFs to communicate with each other. Note that it is not completely transparent to the NFs, as some additional parameters addressed to the SCP are included in the messages, as specified in 3GPP TS 29.500 [9] (e.g., signaling interface in Oracle's SCP solution [10]). For example, when an NF service consumer sends an NF service request to an NF service producer through the SCP, the NF service consumer includes some SCP-specific parameters for discovery and selection (e.g., `3gpp-Sbi-Discovery`). This is also true for indirect communication messages (for example, `3gpp-Sbi-Target-apiRoot`). These parameters are typically included in the requests as 5G-specific HTTP headers.

### C. SCP DEPLOYMENTS

Annex G of TS 23.501 presents three possible SCP deployments as examples: SCP based on service mesh, SCP based on independent deployment units, and SCP based on name-based routing. Fig. 1 shows the architecture of the SCP deployment examples.

- SCP based on service mesh: The architecture consists of SCP agents and an SCP mesh controller. Each NF service is co-located with an SCP agent in the same deployment unit. Here ''deployment unit'' refers to the virtual network function component (VNFC) in which the software is executed (e.g. a container-based VNFC). According to [5], this deployment is expected to become the norm. The architecture of this deployment is shown in Fig. 1a.
- SCP based on independent deployment units: The architecture also consists of SCP agents and an SCP controller. These elements are deployed as units that are independent of NF services. Fig. 1c shows the architecture of this deployment.
- SCP based on name-based routing: The architecture comprises a service router, a path computation element, a discovery service, and a registration service. The service deployment cluster includes NF services and the service router. The SCP platform provides discovery, registration, and routing through a path computation element, and forwarding through an underlying Internet protocol (IP) over an information-centric network (ICN) [11]. The architecture of this deployment is shown in Fig. 1b.

### D. SDN

The SDN [12] paradigm makes networks flexible by decoupling the data plane from the control and management planes. The data plane is executed by SDN switches, which are plain forwarding devices, and the control plane is logically centralized in an entity named the SDN controller. The management plane is executed by SDN applications running on top of the SDN controller.

SDN switches are managed by flow rules, where each flow rule defines a match criterion (i.e. to which packets the flow rule applies) and a set of actions (i.e., what to do with the packets), and maintains a set of counters (e.g., how many packets have been matched). SDN applications define the policies of the network and indicate how the SDN controller configures the flow rules of the switches to achieve the desired network behavior. The centralized view of the control plane allows for flexibility, adaptability, and optimal network configurations.

The important role of SDN in flexible and programmable next-generation cellular networks has been discussed in the literature [6], [7]. Existing proposals include realizing the user plane function (UPF) as an SDN switch [13]–[15], replacing the GTP protocol with MPLS to reduce user plane traffic overhead, and simplified handover proce-dures [16]. SoftCell [17] proposed a clean-slate core network architecture composed of SDN switches and middleboxes that avoids all the standard entities and protocols of the 3GPP.

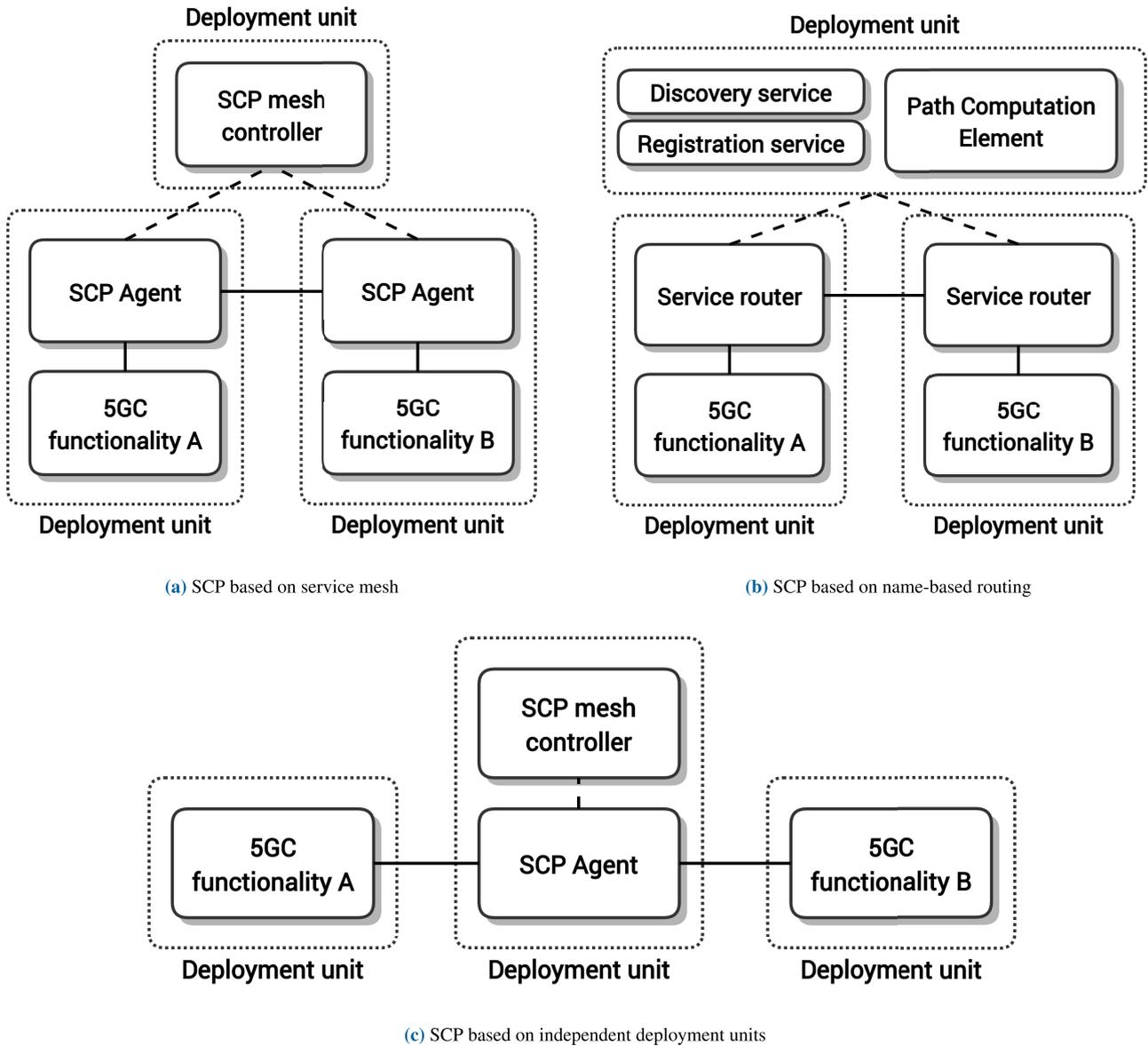

**FIGURE 1.** Architecture of the SCP deployment examples. Adapted from [3].

Network function virtualization (NFV) technology is increasingly being considered along with SDN to increase the flexibility of cellular networks. They are key enablers for the integration of network slicing and edge computing paradigms in 5G networks. In this regard, some works focused on the orchestration of network slicing and edge computing in 5G networks [18], [19]. However, to the best of our knowledge, replacing SCP with an SDN solution has not yet been proposed. Existing SDN routing and forwarding applications focus on the efficient delivery of traffic among different hosts based on network layer information. In this work, we propose a novel SDN application which also considers 5G-specific information (e.g. the load of the NF instances) for providing efficient signaling communications. The proposed SDN application also considers NF endpoint translation for transparently handling the communications between the NFs.

## III. SDN-BASED SOLUTION FOR INTERCONNECTING NETWORK FUNCTIONS

We propose using SDN to build a completely transparent solution as an alternative to the SCP. We leverage the strengths of SDN to dynamically control the internal traffic among different NFs. Fig. 2 shows the proposed architecture. It considers the introduction of SDN switches for connecting the NFs and an SDN controller with SDN applications running on top. Regarding the SDN switches, we consider that either physical or virtual switches can be co-located with each NF. In our proposal, a dedicated SDN application (SCP APP in Fig. 2) is in charge of providing the functionalities of the SCP (e.g., connectivity, service discovery, load balancing, etc.). The dedicated SDN application manages the flow rules installed on the SDN switch of each NF in order to provide the desired behavior. Note that the decisions of SCP applications correspond to the SCP entity, according to operator-defined policies. Thus, the indirect communication capabilities among the NFs are completely transparent to the NFs. At the same time, SCP features such as load balancing, monitoring, and overload control are available.

Unlike the SCP, our SDN-based mechanism does not require NFs to include additional parameters in NF service requests/responses, as the SDN controller handles the policies internally. As a result, our proposed solution is completely transparent to the NFs. It also reduces the complexity introduced by the SCP in the network by eliminating the SCP intermediary and handling the communications directly through SDN. In addition, it has the potential to remove the data plane overhead introduced by SCP agents if the SDN switches can process most packets at the line rate. This is achieved by moving the decision logic from the SCP agents to the SDN controller. Consequently, our proposal can also be interpreted as a mechanism for implementing some of the SCP functionalities transparently to the NFs. This allows 3GPP-compliant SCP implementations to leverage our proposed SDN-based solution for implementing the desired functionalities.

An SDN-based architecture may have some limitations, the main one being that SDN switches are only able to match and perform actions on packet headers up to the transport layer. Thus, by default, they cannot parse application data such as the content of the HTTP/2 application layer protocol adopted in 5G for service-based interfaces.

On the other hand, SDN applications running on top of the SDN controller are software applications that can process received packets as desired. Consequently, there is a trade-off between direct packet processing at the switches, which yields great performance but is subject to the limits of inspection capabilities, and packet forwarding to the SDN controller for further processing, which hinders performance but may enable full packet inspection.

Therefore, our goal is to process as many packets as possible on the switches themselves while minimizing the number of packets that must be sent to the SDN controller. To this end, our solution only forwards to the SDN controller the initial packet of a new communication between two NFs and the packets destined to the NRF. The rest of the traffic is forwarded by the SDN switches at the line rate.

The rest of this section describes how the proposed SDN application provides the different functionalities offered by the SCP. In addition, we analyze the SDN signaling load generated by our solution (i.e., the SDN control plane traffic exchanged between the SDN controller and the SDN switches). However, our solution is not tied to a specific protocol for the southbound interface, we refer to OpenFlow in the analysis, as it is the preferred choice of the SDN community.

### A. COMMUNICATION: MESSAGE FORWARDING AND ROUTING

The underlying SDN network composed of SDN switches provides communication capabilities between the NFs. There is no longer a distinction between ''direct'' and ''indirect'' types of communication. The communication between two NFs does not have to traverse any SCP agent but just the network (in this case, built with SDN-enabled switches). From the viewpoint of NFs, communications are always direct.

The SDN application provides connectivity in reactive forwarding mode: received packets at the SDN switch that do not match any flow rule are sent to the SDN controller to be processed by the SDN application. Then, the SDN controller indicates to the SDN switches how to forward those packets and installs a forwarding flow rule for the packets of that flow at the SDN switch, so that subsequent packets are directly forwarded by the switch without reaching the SDN controller.

This feature will be beneficial not only in terms of improved performance but also in terms of simplified configuration of the NFs compared to the SCP alternative.

This procedure involves some SDN signaling when a new communication is started between two NFs. First, an OpenFlow `PACKET-IN` message encapsulating the first packet that is received is sent from the SDN switch to the SDN controller. Then, the SDN controller sends a `PACKET-OUT` message to the SDN switch indicating the actions to be applied to the packet. In addition, the SDN controller will also send a `FLOW-MOD` message to the SDN switch indicating the flow rule to be installed in the switch for forwarding all subsequent packets of that communication.

### B. DISCOVERY

NF service discovery messages destined to the NRF are forwarded from the SDN switch that is co-located with the NF to the SDN controller. The SDN application implements this by installing a flow rule in the SDN switches, so that all packets destined to the NRF are sent to the SDN controller. Note that the NF service registration and de-registration messages are forwarded transparently to the NRF.

When the SDN application receives an NF service discovery message from an NF service, it submits an NF service discovery request to the NRF only if the SDN application does not have previously stored information about the requested service. Then, the SDN controller answers the request by providing only a single endpoint (IP address and TCP port) representing the main endpoint of the NF service producer type to be discovered.

In this way, each NF service producer type will be represented by a single endpoint, which will be used by all NF service consumers seeking to communicate with the corresponding NF service producer. This also contributes to simplifying the NFs, because, from the NF service consumer point of view, there is only one NF service producer instance of each type, and thus instance selection is trivial. Actual forwarding to the desired NF service producer instance can be achieved by performing a translation of the endpoints at the SDN switches as described in Section III-D.

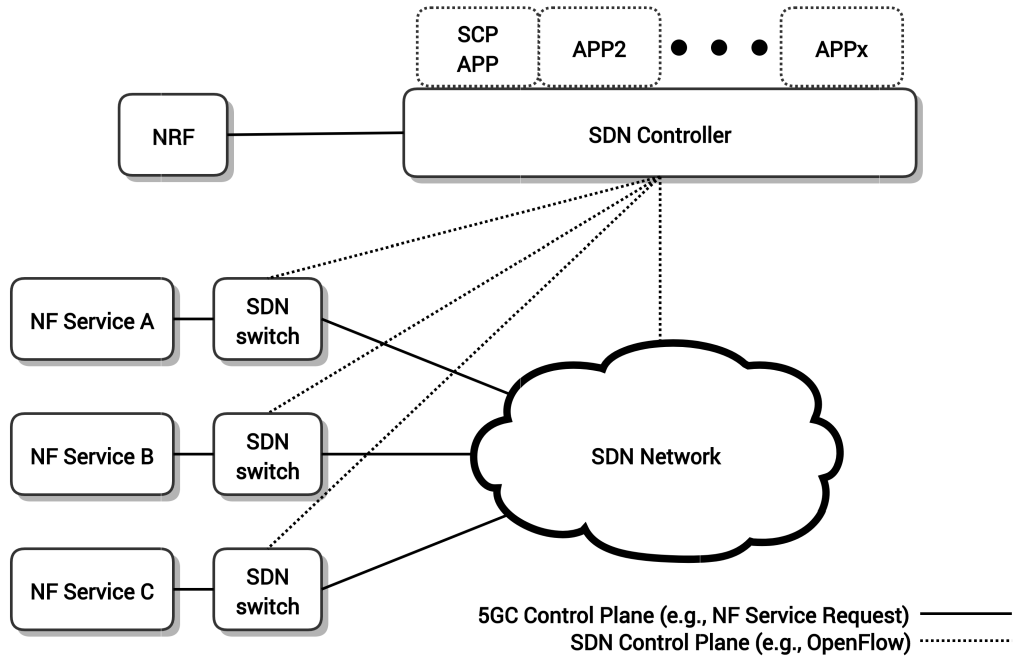

**FIGURE 2.** The architecture of the proposed solution.

Every communication destined to the NRF (e.g., discovery, registration, or de-registration) involves the following SDN signaling. First, an OpenFlow `PACKET-IN` message is sent from the SDN switch to the SDN controller encapsulating the packet destined to the NRF. Then, the SDN controller sends a `PACKET-OUT` message to the SDN switch. In the case of registration or de-registration, this packet will encapsulate the original packet with the instruction to forward the packet to the NRF. In the case of discovery, the encapsulated packet carries the discovery results destined to the consumer NF.

### C. COMMUNICATIONS SECURITY

In terms of security, our proposed solution is analogous to the co-located SCP agent scenario studied by the 3GPP, where security is provided by physical co-location of the SCP agent (SDN switch in our proposal) and the NFs as stated in 3GPP TS 33.501 [20]. In addition, the packets that are sent from the SDN switch to the SDN application are forwarded through the secure control channel established between the SDN switch and the SDN controller (e.g., transport layer security (TLS) channel in OpenFlow).

Communications security also encompasses validating the authorization of the NF service consumer to access a given NF service producer type. This should be implemented both in the service discovery phase and during communication between NF services.

In the discovery phase, when the SDN application receives an NF service discovery request, it checks that the requesting NF is allowed to access the requested NF service producer. If not, the SDN application replies to the NF service consumer with the proper unauthorized error, and the discovery process does not continue.

During communication between NF services, when an NF service consumer tries to access a given NF service producer, the first packet of a new communication is sent to the SDN application as described in the reactive forwarding in Section III-A. Thus, upon receiving the packet, the SDN application checks whether the requesting NF is authorized to access the desired NF service producer. If not, the SDN application replies to the NF service consumer with an unauthorized error and no flow rules are installed on the SDN switch.

Note that this procedure does not add further SDN signaling to that described in the previous subsections.

### D. LOAD BALANCING

A particular NF service producer instance can be selected by adequately configuring the flow rules of the SDN switch that is co-located with the NF service consumer. Note that the requests of the NF service consumer instance will be addressed to the main endpoint of the NF service producer type, which is provided to the NF service consumer by the SDN application at discovery time. Thus, by mapping the main endpoint to the real destination address at the SDN switch, requests can be redirected to any of the instances that provide the desired service.

To this end, the SDN application achieves this behavior by only configuring two flow rules at the SDN switches that are co-located with NF service consumers:

- One rule matches the destination IP address and TCP port of the main endpoint of the NF service producer that replaces the IP address with the address of the endpoint of one of the NF service producer instances (based on the defined load balancing policy).
- Another rule matches the source IP address and TCP port of the endpoint of the actual NF service producer instance, which replaces them with those of the main endpoint of the NF service producer.

This operation is compatible with IPv4 and IPv6 network addresses, since the OpenFlow protocol directly supports matching and changing for these addresses since version 1.2. Also note that this procedure does not require the NF instances to perform any operation. NF service producer instances can be dynamically scaled, and NF service consumer instances do not require any modification. They issue their requests to the main endpoint address, and the flow rules configured in their SDN switches will redirect the messages to the desired instance. We also remark that our proposal does not enforce any specific load-balancing policy. Instead, state-of-the-art algorithms for making load balancing decisions (e.g., [21]–[23]) can be integrated on top of our SDN-based solution. The load-balancing algorithm is implemented as part of the SDN application running on top of the SDN controller.

### E. MONITORING AND OVERLOAD CONTROL

Similar to the SCP agent, the proposed SDN application monitors the status and load of the NFs (for example, CPU consumption, memory, and disk usage). The SDN application can obtain these from the NF profile stored in the NRF, which contains information about the NF load and capacity, as described in 3GPP TS 23.501 [3].

In addition, from a network point of view, NF services can also be monitored with our solution by analyzing the counters of the flow rules of the switches, which can provide the number of packets and bytes that match the rules and the elapsed times since the rules were installed.

These two complementary sources of information can be used to balance the load between the different NF instances. Furthermore, they can be used as an access control mechanism for denying NF service requests if all NF service producers are saturated, thereby preventing overload situations.

This procedure does not introduce additional SDN signaling into the solution, but leverages the existing SDN switch monitoring by the SDN controller. The SDN switches send to the SDN controller flow statistics included in the OpenFlow `MULTIPART` messages. These messages are asynchronously transmitted either when a threshold is reached (e.g., number of bytes in a flow) or periodically. This relieves the SDN controller from continuously polling SDN switch counters, thus reducing the SDN control plane overhead.

### F. ROUTING BINDING

Routing binding refers to explicitly establishing a communication channel between an NF service consumer instance and an NF service producer instance, to be used by subsequent messages.

In the proposed solution, routing binding is intrinsic because flow rules are established in the SDN switches for the duration of a communication between an NF service consumer and an NF service producer.

The challenge lies in fulfilling the routing binding indications included in NF service consumer requests if dynamic load balancing is to be provided (i.e., changing the NF service producer instance assigned to an NF service consumer during an active communication). Note, however, that the routing binding indications contained in NF service producer notifications can always be satisfied because they will be destined to a specific consumer.

To support routing binding under dynamic load balancing, we propose setting and idle timeout counter. That is, after a given inactivity period in the communication between an NF service consumer and an NF service producer, the corresponding flow rules in the SDN switches are removed. Then, the next received packet will be sent to the SDN controller. The SDN controller can then determine whether the routing binding is required or, instead, if the NF Producer instance can be chosen based on a load-balancing policy. This guarantees that most routing binding indications will be satisfied and that transactions will be successfully completed while simultaneously allowing for dynamic load balancing.

OpenFlow directly supports the capability to remove flow rules after idle timeouts when it installs the flow rules in the SDN switch. Thus, no further signaling is required.

### G. SCALABILITY

mMTC scenarios require handling a large number of UEs connected to the network. This has direct a impact on the 5GC signaling load (i.e., the number of messages exchanged between control plane NFs) [2]. However, the impact in the traffic that must pass through the SDN application (i.e., SDN signaling load) is several orders of magnitude lower and thus negligible. This is because a single NF can serve a large number of UEs. In addition, note that a growing number of UEs has no considerable impact on the amount of traffic exchanged with the NRF (e.g., NF service discoveries or registrations) and consequently the performance of our solution is preserved.

In summary, increasing the number of UEs connected to the network results in a considerable increase in the 5GC signaling load but not in the SDN signaling load. Both the SDN signaling load and the sizes of switch flow tables depend on the number of active NFs, but not on the level of the 5GC

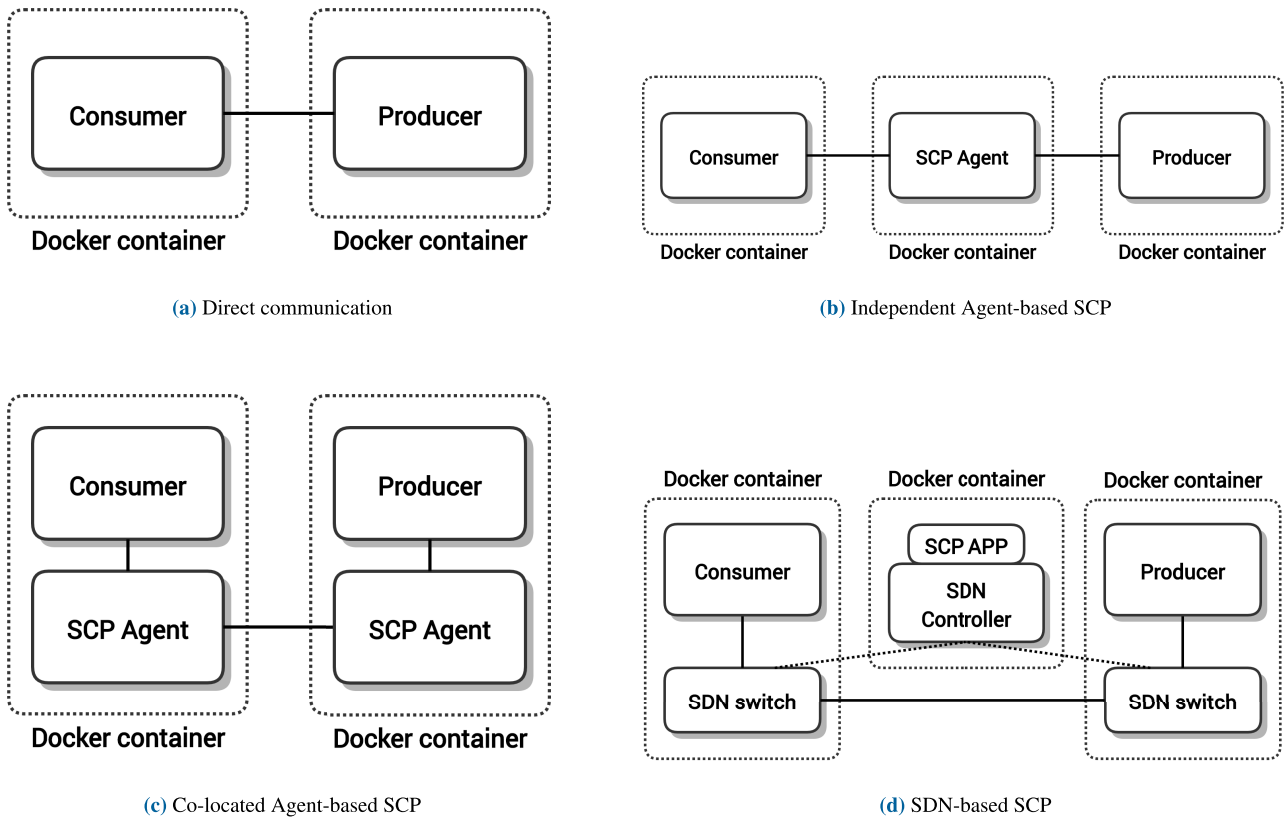

**FIGURE 3.** Setups for performance evaluation in the experiments.

signaling load. As a result, our proposed solution scales well in mMTC scenarios.

## IV. ANALYSIS
In this section, we analyze our proposed SDN-based solution and compare it with a direct communication scenario and two agent-based SCP solutions: one based on independent deployment units and another on co-located deployments.

### A. PERFORMANCE EVALUATION
Experiments were conducted to evaluate the throughput and latency of the proposed solution. NFs were deployed in Docker containers running on commodity hardware (Intel® Core™ i9-9900K CPU @3.60GHz). We allocated two dedicated CPUs to each Docker container.

Specifically, we measured the communication performance between a consumer NF and a producer NF. The producer NF was implemented using Spring Boot [24], which provides a state-of-the-art framework for creating REST services. The producer NF exposed an HTTP endpoint that returned a random string of the desired length. For the consumer NF, we used the `wrk` [25] open-source HTTP benchmarking tool, which allows the evaluation of the performance of production-grade HTTP APIs by maintaining multiple parallel connections between the client and the server. We used the HTTP/1.1 version because at the moment this paper was written, `wrk` did not support HTTP/2. This tool generates as many HTTP requests as possible, thus saturating the connections that are established with the HTTP server. This allows to determine the maximum throughput supported by the communication channel between the consumer and the producer. Alternatively, we have also used the `vegeta` [26] open-source HTTP load testing tool, which allows to generate HTTP requests at constant rate. This allows to measure the latency of the communication channel between the consumer and the producer when it is not saturated.

Fig. 3 depicts the setups in the different communication scenarios in the analysis.

In the direct communication scenario (Fig. 3a), the consumer NF interacted directly with the endpoint exposed by the producer. This is a simpler setup that does not address the main challenges of a service-mesh architecture. Nevertheless, it allows the establishment of a reference baseline with the maximum performance.

For the agent-based SCP scenario, we implemented the SCP agent using the Spring Boot framework. The SCP agent acted as a simple proxy exposing an HTTP endpoint that received the target endpoint as a parameter (that is, producer NF or another SCP agent), by requesting the target on behalf of the client, gathering the response, and returning the response to the client. In the case of independent SCP deployment (Fig. 3b), an SCP agent was deployed in a Docker container. In the case of co-located agents (Fig. 3c), SCP agents were introduced inside the consumer and producer NF Docker containers.

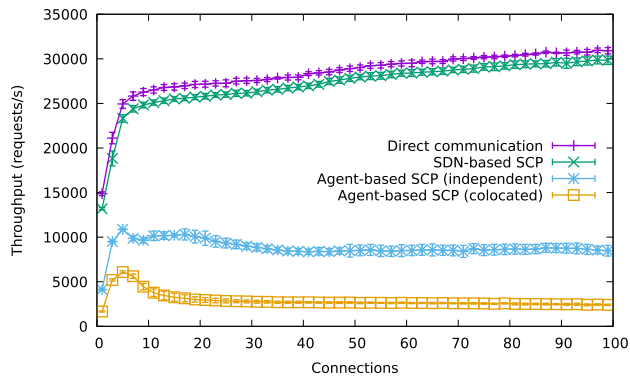

**FIGURE 4.** Throughput versus the number of connections using the `wrk` tool.

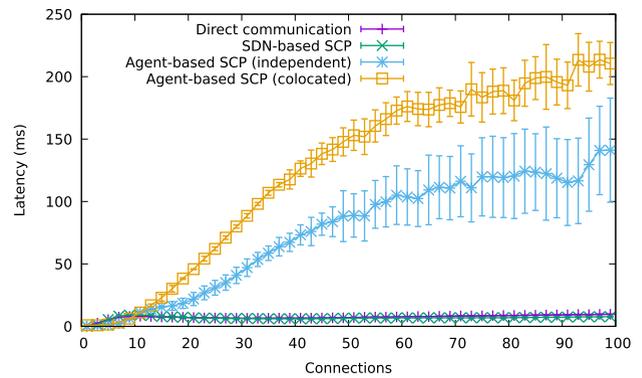

**FIGURE 5.** Latency (95th percentile) versus the number of connections using the `wrk` tool.

For the scenario with our SDN-based proposal (Fig. 3d), we implemented a testbed with different open-source SDN tools: Open vSwitch (OVS) [27] for the SDN switches that were co-located with the NFs and Open Network Operating System (ONOS) [28] for the SDN controller. We implemented an SDN application on top of the SDN controller to manage the flow tables of the SDN switches providing connectivity between the two NFs. The SDN controller was executed in a Docker container. SDN switches were introduced inside the consumer and producer Docker containers.

Regarding the network configuration, we used a dedicated bridge network to connect the containers, and assigned IP addresses in the same subnet. In case of co-located SCP agents, different TCP ports were used for the SCP agent and the NF in each container. Besides, in our SDN-based proposal, the SDN switch was configured with the IP address of the container, while the main network interface of the container was added as a port of the SDN switch.

To calculate the performance of the different alternatives under a saturation regime, we used the `wrk` tool to measure the throughput and latency in the consumer NF. Each test lasted 60 seconds, by varying the number of parallel connections from 1 to 99 in steps of 2. The throughput was measured as the rate of HTTP requests per second which that the consumer NF can send and the producer NF can respond within a timeout of 30 seconds. The latency was measured as the round-trip time from sending the HTTP request to receiving the HTTP response at the consumer NF. To calculate the latency in scenarios without saturation, we used the `vegeta` tool. Each test lasted 60 seconds, by varying the rate of generated HTTP requests from 100 to 1000 requests per second in steps of 100. The results correspond to an average of 10 independent executions with 95 % confidence intervals.

Fig. 4 shows the throughput versus the number of connections in the saturated regime test using the `wrk` tool. The highest throughput was attained in the direct communication scenario. This was expected because, in this scenario, the communications between the consumer and the producer did not pass through any intermediate entity. The throughput grew rapidly up to 25 000 requests/s for five connections. Then, it increased slowly up to 31 000 requests/s for 100 connections.

The throughput levels in the SDN-based scenario were very similar to those in the direct communication scenario. The trend was similar, with a gap of approximately 1000 requests/s. This means a sub-5 % overhead of above 20 000 requests/s. As a result, the SDN-based solution could achieve up to 30 000 requests/s in the experiment.

The agent-based SCP solutions became saturated much earlier. Specifically, the independent deployment became saturated at 10 000 requests/s, whereas the co-located deployment could not serve more than 5000 requests/s. The approximate overheads that these values represent are 65 % and 84 %, respectively, compared to direct communication. This significant decrease in performance was mainly caused by the processing tasks at the SCP agent to parse the content of the HTTP headers. Note that in the co-located SCP agents scenario, the achievable throughput was considerably less than that in the independent deployment scenario. This is directly related to the fact that, in the former, packets traverse two SCP agents deployed in the same container as the NF services (one in the consumer and another in the producer). Thus, SCP agents and NF services share the same set of resources, which results in CPU saturation for a noticeably lower throughput than in the independent scenario, where packets traverse a single SCP agent that is deployed in a dedicated container.

Fig. 5 shows 95th percentile latency results versus number of connections in the saturated regime tests using the `wrk` tool.

We can observe that the latency values for the direct communication and SDN-based scenarios were very stable below 10 ms for the 95th percentile regardless of the number of connections. However, this was not the case for agent-based SCP scenarios, where latency increased linearly with the number of connections. Again, the scenario with the independent deployment of SCP agents achieved better results than their co-location, by yielding 95th percentile latency values of approximately 50 ms for 30 connections, compared to 80 ms in the second case. For 60 connections, the respective values were approximately 100 ms and 160 ms, one order of magnitude higher than those in the direct communication and SDN-based scenarios.

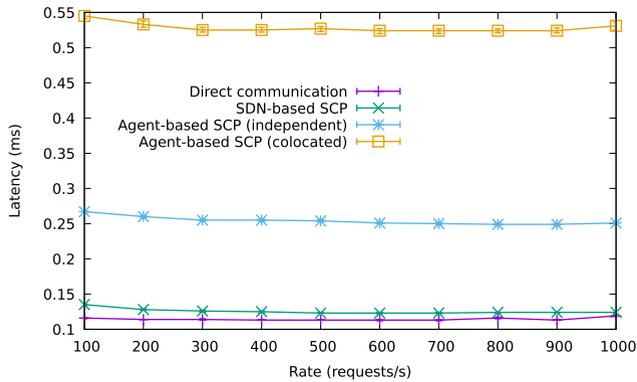

**FIGURE 6.** Latency (95th percentile) versus rate of requests using the `vegeta` tool.

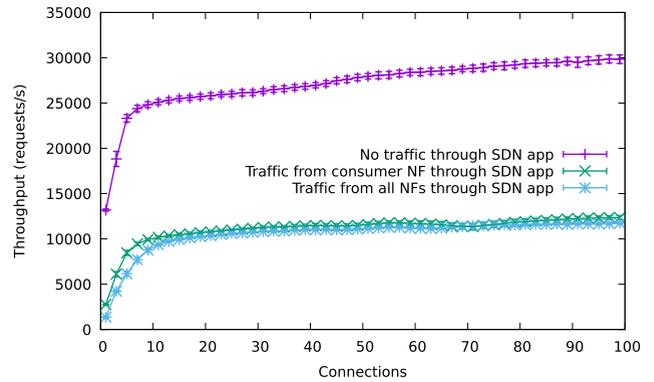

**FIGURE 7.** Throughput for different SDN-based setups using the `wrk` tool.

Fig. 6 shows the 95th percentile latency versus the rate of HTTP requests using the `vegeta` tool. We can observe that the latency levels attained by the SDN-based proposal are almost identical to those obtained in the direct communication setup baseline. The independent agent-based SCP doubles these levels, and the co-located deployment quadruples those of the baseline. Overall, these results confirm that the overhead introduced by our proposed SDN solution in terms of latency are negligible. On the other hand, agent-based implementations of the SCP introduce a delay in the communications even in scenarios without congestion.

### B. TRAFFIC FORWARDED THROUGH THE SDN APPLICATION

In this section, we analyze the impact of traffic that has to traverse the SDN application with our solution. We used the same setup and methodology as in the previous experiment; however, in this case we compared three modified versions of the SDN application:

- No traffic is forwarded through the SDN application. The SDN application proactively configures the SDN switches with flow rules to forward the packets to the desired port.
- The traffic at the SDN switch of the NF consumer is forwarded through the SDN application.
- The traffic at the SDN switches of the consumer and producer NFs is forwarded through the SDN application.

Fig. 7 shows the results. They reveal the performance degradation due to traffic forwarding through the SDN application running on top of the SDN controller. We can observe that, when traffic is forwarded through the NF application, the maximum achieved throughput is close to 11 000 requests/s, which means an overhead of approximately 19 000 requests/s compared to direct traffic forwarding at the SDN switches without traversing the SDN controller. Interestingly, we can observe that there is no significant performance penalty for sending the traffic of the two NFs through the SDN application compared to forwarding the traffic of just one of them because both cases exceed 10 000 requests/s. These results are in line with the agent-based SCP scenario with independent deployment, where the packets are processed by an intermediary in a dedicated container.

Moreover, the results highlight the importance of minimizing the number of packets that must be sent through the SDN controller to optimize the performance. Recall that in our solution packets need to traverse the SDN application either when a communication is destined to the NRF or when there is no specific flow rule for that packet in an NF SDN switch (e.g., it is the first packet of a communication between two NFs). According to 3GPP TS 23.501 [3], consumer NFs cache the result from valid NF discoveries for a given validity period and reuse that information for subsequent communications. This contributes to reducing the number of triggered NF service discovery actions, which improves the performance of our proposed solution. Furthermore, consumer NFs can set very long validity periods for discovery results or even store them permanently since the SDN application will update target NF producer instances as mandated by the load-balancing policies.

As a result, communications with the NRF do not benefit from our solution because they are always forwarded through the SDN application. Nevertheless, these communications represent a small fraction of the 5GC control plane traffic.

We conducted some simulations to determine the amount of 5GC control-plane traffic that must be forwarded through the SDN application in our proposal. To this end, we consider the communications between NFs related to the attachment and detachment procedures as in [29]. This involves the communication between the NRF, AMF, SMF and the Authentication Server Function (AUSF). We consider that each NF registers its services at the NRF at the beginning. Then, a service discovery is triggered whenever an NF needs to communicate with another NF service. As depicted in the call flow diagram in [29], we consider that the attachment procedure triggers the exchange of four messages between the AMF and the AUSF (Authentication Request/Reply and Location Update Request/Reply) and four messages between the AMF and the SMF (Session Request/Response and Modify Bearer Request/Response). The detachment procedure only involves two messages between the AMF and the SMF (Detach Request/Response).

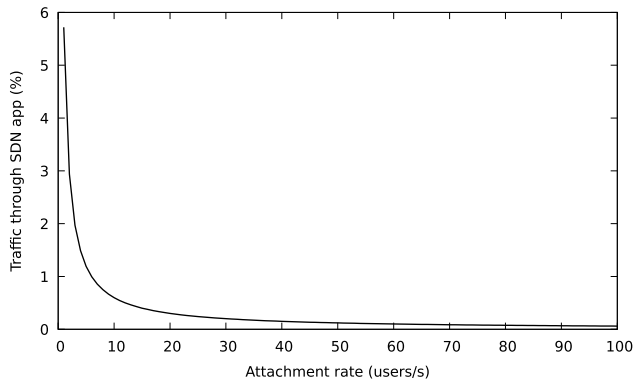

**FIGURE 8.** Percentage of packets that are forwarded through the SDN application as the user attachment rate increases.

For the simulation, we consider that the service discovery results are cached at the NFs for a given validity period of 10 seconds and a flow rule hard timeout of 20 seconds. We consider a scenario for our proposal in which users get attached to the network at a constant rate and remain attached for a fixed timespan of 1 minute.

Fig. 8 shows the percentage of packets that are forwarded through the SDN application as the attachment rate increases during a simulation interval of 1 hour. As we can observe, the percentage of packets that are sent through the SDN controller decreases exponentially as the user attachment rate increases, being lower than 6 % for a one user attachment per second. The value drops below 1 % for an attachment rate of 6 users per second. The results of this simulation highlight that the proposed solution scales well as the 5GC signaling load increases, and that it can handle most control-plane traffic at the SDN switches.

## V. CONCLUSION

The SBA model recommended for new 5G network cores allows the implementation of functionalities as fine-grained microservices that decouple the services from the infrastructure. This makes it possible to build flexible architectures that can be distributed in a cloud environment with scaling and fail-over mechanisms. These architectures can adapt the core in real time to traffic loads or other dynamic requirements. Nevertheless, the variability of the 5G core is also challenging. Therefore, the 3GPP has recently introduced the SCP entity to abstract control plane signaling communications in the 5GC SBA. This entity provides indirect communication and delegated discovery capabilities to the NFs. However, it is not completely transparent to the NFs and therefore adds complexity and overhead.

This paper proposes a completely transparent SDN-based alternative to the SCP for interconnecting network functions. This novel solution addresses the same issues as the SCP, but it is completely transparent to the NFs, which simplifies their implementation.

We evaluated the performance of our solution in terms of control plane throughput and latency and compared it with direct communication and agent-based SCP implementations. Our experimental results show that the performance of an SDN-based implementation is in line with the direct communication between NFs (the simplest alternative, which does not address any of the goals of the SBA architecture), whereas agent-based SCP implementations may introduce substantial overhead as the signaling load increases.

From our experiments, we can conclude that by relying on SDN technology, it is possible to achieve the functionalities of the SCP without introducing any additional entity in the 5GC.

In addition to eliminating the SCP, our solution also simplifies the NFs of the 5GC because they do not have to be configured for direct or indirect signaling communication, nor include any additional parameters in NF service requests/responses.

Overall, the proposed SDN-based solution has the benefits of SCP in terms of communication handling between NF services, with a performance close to that of direct communication. We are aware of the limitations that SDN imposes on the functionalities that can be provided because the packets cannot be inspected in the SDN switches in depth. However, our solution circumvents this limitation, while supporting all the basic features expected from the SCP, by sending selected packets through the SDN application. Moreover, our results reveal that the percentage of packets that must be sent through the SDN application decreases with the 5GC signaling load.

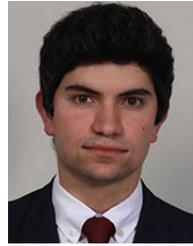

**PABLO FONDO-FERREIRO** received the bachelor's and master's degrees in telecommunication engineering from the University of Vigo, in 2016 and 2018, respectively, where he is currently pursuing the Ph.D. degree. His research interests include SDN, mobile networks, and artificial intelligence. He received the Award for the Best Academic Record from the University of Vigo. In 2016, he received a Collaboration Grant from the Spanish Ministry of Education for Research in SDN and Energy Efficiency in Communication Networks. In 2018, he received a Fellowship from "la Caixa" Foundation to pursue his Ph.D. degree at the University of Vigo.

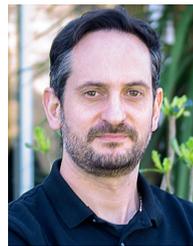

**FELIPE GIL-CASTIÑEIRA** received the M.Sc. and Ph.D. degrees in telecommunication engineering from the University of Vigo, in 2002 and 2007, respectively. He is currently an Associate Professor with the Department of Telematics Engineering, University of Vigo. Between 2014 and September 2016, he was the Head of the iNetS area with the Galician Research and Development Center in Advanced Telecommunications. He has published over 60 papers in international journals and conference proceedings. He has led several national and international research and development projects. He holds two patents in mobile communications. He is also a Co-Founder of a university spin-off, Ancora. His research interests include wireless communication and core network technologies, multimedia communications, embedded systems, ubiquitous computing, and the Internet of Things.

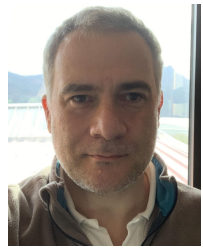

**FRANCISCO JAVIER GONZÁLEZ-CASTAÑO** is currently a Catedrático de Universidad (a Full Professor) with the Telematics Engineering Department, University of Vigo, Spain, where he leads the Information Technology Group. He has authored over 100 articles in international journals in the fields of telecommunications and computer science, and has participated in several relevant national and international projects. He holds three U.S. patents.

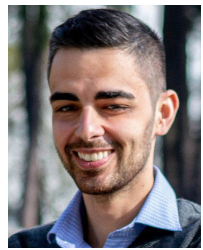

**DAVID CANDAL-VENTUREIRA** received the bachelor's and master's degrees in telecommunication engineering from the University of Vigo, in 2016 and 2018, respectively. He is currently pursuing the Ph.D. degree in information and communication technologies. Since 2018, he has been working as a Researcher with the Information Technologies Group, University of Vigo. His research interests include mobile and wireless networks and artificial intelligence.

◦ ◦ ◦